\documentclass[aps,prd,onecolumn,groupedaddress,showpacs]{revtex4}
\usepackage{epsfig} 
\usepackage{graphics}
\usepackage{amsfonts}

\begin{document}

\title{Walls and Chains of Planar Skyrmions}
\author{Derek Harland}\email{d.g.harland@durham.ac.uk}
  \author{R. S. Ward}\email{richard.ward@durham.ac.uk}
\affiliation{Department of Mathematical Sciences, University of
Durham, Durham DH1 3LE}

\date{\today}

\begin{abstract}
In planar (baby) Skyrme systems, there may be extended linear structures
which resemble either domain walls or chains of skyrmions, depending
on the choice of potential and boundary conditions. We show that
systems with a single vacuum, for example with potential $V=1-\phi_3$,
admit chain solutions, whereas walls are
ruled out by the uniqueness of the vacuum. On the other hand, in
double-vacuum systems such as $V=\frac{1}{2}(1-\phi_3^2)$, one has
stable wall solutions, but there are no stable chains; the walls may be
viewed as the primary objects in such systems, with skyrmions being
made out of them.
\end{abstract}


\maketitle

\section{Introduction}

The (2+1)-dimensional Skyrme system (known as the planar Skyrme model, or
baby Skyrme model) may be viewed as a simple analog of the
(3+1)-dimensional Skyrme model, and is also of considerable interest 
in its own right. This paper is concerned with walls and chains of
planar skyrmions. Walls are extended objects of
codimension 1, and chains are extended objects of dimension 1; so in the
planar Skyrme system, walls and chains are both 1-dimensional objects,
and they are related, as we shall see below.

In addition to the usual isolated (0-dimensional) skyrmion solutions,
various other periodic, extended solutions are known for both the Skyrme
and the baby Skyrme models: for example, a 3-dimensional skyrmion crystal
\cite{klebanov85}, a 2-dimensional skyrmion wall \cite{battye&sutcliffe98},
and a planar skyrmion crystal \cite{ward04}. But 1-dimensional chains of
skyrmions, or of planar skyrmions, have received less attention.

Chains of topological solitons have been intensively studied in other systems:
for example, chains of instantons (`calorons') in the O(3) sigma model
\cite{mottola89} and in Yang-Mills theory \cite{HS78}, and chains of
BPS monopoles \cite{CK01, ward05}. A chain may be viewed as a string of
solitons, and the dimensionless ratio
$C$=(intersoliton distance)/(soliton size) is an important parameter.
When $C<1$, interesting things can happen: 
the solitons lose their individual identities, and `constituents' may
emerge. In the caloron case, the constituents are monopoles
\cite{kraan&vanbaal98,lee&lu98,bruck07}.

One fundamental difference between chains of skyrmions or baby skyrmions on
the one hand, and chains of instantons or monopoles on the other hand, is
as follows. Instantons and monopoles do not exert forces on each other,
and so the energy-per-period of a chain is independent of the parameter $C$.
By contrast, skyrmions and baby skyrmions may attract or repel each other,
depending on their relative orientation; and so we expect a chain to contract,
extend, or exhibit other dynamical properties. Of particular interest are
stable chains: that is, chains whose energy cannot be lowered by any periodic
deformation. Chains of this type might serve as good
approximations to finite chains in $\mathbb{R}^2$ or $\mathbb{R}^3$.

An outline of the rest of this paper is as follows. In section 2
we review the baby Skyrme model and its topology; in sections 3 and 4
we discuss planar skyrmion walls and chains repectively; and section 5 contains
some concluding remarks.


\section{Planar Skyrmions}   

A planar Skyrme (or baby Skyrme) field is a smooth function $\phi$ from
(2+1)-dimensional Minkowski
space-time to the 2-sphere $S^2$. In what follows, we shall restrict to static
fields, and think of $\phi$ as a unit vector $\vec\phi=(\phi_1,\phi_2,\phi_3)$; so 
$\vec\phi=\vec\phi(x,y)$ with $\vec\phi\cdot\vec\phi=1$. Here $x$ and $y$ are the
usual spatial coordinates $x^j=(x^1,x^2)=(x,y)$. The (static) energy density
$\mathcal{E}$ of the field is defined by
\begin{equation}
\label{energy density}
 \mathcal{E}[\phi] = \mathcal{E}_2[\phi] 
      + \mu\mathcal{E}_4[\phi] +\mu\mathcal{E}_0[\phi],
\end{equation}
where
\begin{equation}
\label{energy density parts}
 \mathcal{E}_2 = \frac{1}{2}(\partial_j\phi\cdot\partial_j\phi), \quad
 \mathcal{E}_4 = \frac{1}{2} (\partial_1\phi \times \partial_2\phi)^2, \quad
 \mathcal{E}_0 = V(\phi),
\end{equation}
and where $\mu$ is a positive constant. (Taking the coefficients of 
$\mathcal{E}_4$ and $\mathcal{E}_0$ to be equal fixes the length-scale.)
Two commonly-studied choices for the potential $V$ are:
\begin{itemize}
 \item $V(\vec\phi) = (1-\phi_3)$ (the `old baby Skyrme model', see eg
          \cite{piette&al95})
 \item $V(\vec\phi) = \frac{1}{2} (1-\phi_3^2)$ (the `new baby Skyrme model',
                      see eg \cite{kudryavtsev&al98,weidig99})
\end{itemize}
The energy densities $\mathcal{E}_2$ and $\mathcal{E}_4$ are invariant
under global SO(3) rotations of the target space $S^2$, but
the choice of potential $V$ breaks this symmetry; in the old and new baby Skyrme
models, the symmetry group is broken to SO(2).

The energy of a Skyrme field $\phi$ is
\begin{equation}
   E[\phi] = \frac{1}{4\pi} \int_M \mathcal{E}[\phi]\,dx^1\,dx^2.
\end{equation}
(The factor of $1/4\pi$ is for convenience.)
Of particular interest are stable skyrmions, namely finite-energy Skyrme fields
which are local minima of $E$. For isolated skyrmions on $\mathbb{R}^2$, finiteness
of energy requires that $\vec\phi$ tends to a constant as
$x^2+y^2\rightarrow\infty$. It follows that $\vec\phi$ can be extended continuously
to a map from $S^2$ to $S^2$, and it therefore has a degree (winding number),
called the topological charge $N\in\mathbb{Z}$. There is an integral formula
for $N$, namely $N[\phi]=\int_{\mathbb{R}^2} \mathcal{N}[\phi] \, dx \, dy$, where
\begin{equation}\label{charge density}
\mathcal{N}[\phi] = \frac{1}{4\pi} \phi\cdot(\partial_1\phi\times\partial_2\phi)
\end{equation}
is the topological charge density. A standard Bogomolny-type argument shows
that the topology gives a lower bound on the energy, namely
\begin{equation}
\label{bogomolny bound}
E \geq N \left( 1 + \sqrt{2}\mu\int_{S^2} \sqrt{V}\,d\omega \right),
\end{equation}
where $d\omega$ denotes the standard integration measure on $S^2$.

In each of the old and the new baby Skyrme models, there exist isolated $N$-Skyrmion
solutions. For $N=1,2$ in the old baby Skyrme model, and for all $N$ in the
new baby Skyrme model, these solutions are rotationally-symmetric: a
spatial $O(2)$ rotation can be compensated by a rotation of the target 2-sphere.
In fact, these solutions have the O(2)-symmetric `hedgehog' form
\begin{equation}
\label{hedgehog}
\phi = ( \sin f(r) \cos(\theta-\chi),\, \sin f(r) \sin(\theta-\chi), \cos f(r) ),
\end{equation}
where $(r,\theta)$ are polar coordinates on $\mathbb{R}^2$, $\chi\in[0,2\pi)$ is 
a phase angle, and $f(r)$ is a function satisfying $f(0)=\pi$ and $f(r)\rightarrow0$
as $r\rightarrow\infty$. The force between separated skyrmions depends partly on
their relative phase.

In the sections that follow, we shall impose periodicity (or anti-periodicity)
in $y$, so that the field $\phi$ lives on $\mathbb{R}\times S^1$ rather than
$\mathbb{R}^2$.
By periodic with period $\beta$ we mean that $\vec\phi(x,y+\beta)=\vec\phi(x,y)$;
while by anti-periodic we mean that $\vec\phi(x,y+\beta)=\sigma\vec\phi(x,y)$,
where $\sigma$ is some element of the global symmetry group SO(2) satisfying
$\sigma^2=1$. The topological classification and Bogomolny bound extend to these
cases, as we shall see below.

For some choices of potential $V$, the Bogomolny bound (\ref{bogomolny bound})
can be saturated, and the corresponding solutions can be written down explicitly.
The procedure for deriving such potentials is well-known
\cite{piette&zakrzewski95,ward04}, and the details may be summarized as follows.
It is convenient to work in terms of the
stereographic projection $W$ of $\vec\phi$, given by
\begin{equation}\label{stereographic projection}
    W=\frac{\phi_1+i\phi_2}{1+\phi_3}.
\end{equation}
In addition, we use the complex coordinate $z=x+iy$. Then the expressions for the
topological charge density (\ref{charge density}) and the energy density
(\ref{energy density parts}) become
\begin{eqnarray*}
\mathcal{N} &=& \frac{1}{\pi} \frac{|\partial_z W|^2-|\partial_{\bar{z}}W|^2}
                       {(|W|^2+1)^2} \\
\mathcal{E}_2 &=& 4 \left( \frac{|\partial_z W|^2+|\partial_{\bar{z}}W|^2}
                       {(|W|^2+1)^2}\right) \\
\mathcal{E}_4 &=& 8 \pi^2 \mathcal{N}^2.
\end{eqnarray*}
The existence of the Bogomolny bound (\ref{bogomolny bound}) follows from the
identities
\begin{eqnarray*}
\mathcal{E}_2 &=& 4\pi\mathcal{N}+\frac{8|\partial_{\bar{z}}W|^2}{(|W|^2+1)^2}, \\
\mathcal{E}_4+\mathcal{E}_0 &=& \left( 2\pi\sqrt{2} \mathcal{N} - 
                 \sqrt{V} \right)^2 + 4\pi\sqrt{2} \mathcal{N} \sqrt{V},
\end{eqnarray*}
and the bound is saturated provided that
\begin{eqnarray}
\label{sigma model bogomolny equation}
\partial_{\bar{z}} W &=& 0 \\
\label{sigma model potential}
V(W) &=& \frac{8 |\partial_z W|^4}{(|W|^2+1)^4}.
\end{eqnarray}
From (\ref{sigma model bogomolny equation}) we see that $W$ has to be a holomorphic
(or rather meromorphic) function of $z$ --- this corresponds to $W$ being a solution
of the $\mu=0$ limit of the system, namely the O(3) sigma model. If we have such
a $W(z)$, and we can express $\partial_z W$ in terms of $W$, then 
(\ref{sigma model potential}) gives the relevant potential $V$. Examples of walls
and chains of this type will be constructed below.


\section{Skyrmion Walls}

In this section, we consider fields which are periodic in $y$ (with period
$\beta$), and which satisfy the boundary condition $\phi_3\to\pm1$ as
$x\to\pm\infty$. In order to get finite energy, the potential $V$ has to
vanish for $\phi_3=\pm1$; the standard example is therefore the
system $V=\frac{1}{2}(1-\phi_3^2)$. The topological classification
is straightforward, and may be described as follows.
The field $\vec\phi$ is defined on the cylinder $\mathbb{R}\times S^1$, but
the boundary condition means that the `ends' at $x=\pm\infty$ can be regarded
as single points, and therefore $\vec\phi$ represents a map from $S^2$ to $S^2$.
So it has a degree $K\in\mathbb{Z}$. Thus the topological (or magnetic)
charge of a segment of wall of length $N\beta$ equals $NK$; note that
the examples below all have $K=1$.
Planar skyrmion walls have been studied previously: for example, the authors of
\cite{kudryavtsev&al98} investigated the interaction between a skyrmion
and an uncharged ($K=0$) wall.

The energy per period $E'$, and the integer $K$, are given by
\begin{eqnarray} 
 E'[\phi] &=& \frac{1}{4\pi}\int_{0}^\beta \int_{-\infty}^{\infty} \mathcal{E}[\phi]
                \,dx^1\,dx^2 \\
 K[\phi] &=& \int_{0}^\beta \int_{-\infty}^{\infty} \mathcal{N}[\phi] \,dx^1\,dx^2,
\end{eqnarray}
where $\mathcal{E}$ and $\mathcal{N}$ are defined by the same expressions
(\ref{energy density}, \ref{charge density}) as before. Furthermore,
the energy $E'$ is bounded below as in (\ref{bogomolny bound}).

One can construct exact solutions by using the procedure described at the
end of the previous section. For simplicity, set $\beta=2\pi$. The simplest
function $W(z)$ satisfying the boundary conditions is $W=\exp(-z)$; substituting
this into (\ref{sigma model potential}), and converting from $W$ to $\phi$,
gives the potential
\begin{equation}
\label{domain wall potential}
V = \frac{1}{2}(1-\phi_3^2)^2. 
\end{equation}
For this potential, the Bogomolny bound is $E'\geq1+\frac{2}{3}\mu$; the
field $W=\exp(-z)$ is a $K=1$ wall solution which saturates this lower bound.
In particular, the wall has a preferred length-per-charge (or period), which
in this case is $2\pi$. (Multiplying $V$ by a constant has the effect of
changing $\beta$.) Note that the size of an individual soliton is of order 1,
so $\beta$ is essentially the dimensionless ratio $C$ mentioned in the Introduction.

Staying with this potential (\ref{domain wall potential}), one can envisage taking
a segment of the wall of length $2\pi N$ in the $y$-direction, and joining the ends
together to form an $N$-skyrmion in $\mathbb{R}^2$. So one might expect $N$-skyrmions
in this system to resemble rings of radius $N$, and with energy
\[
  E = N\left( 1+\frac{2}{3}\mu+\frac{c}{N^2} \right),
\]
where $c$ is a positive constant (the $c/N^2$ term representing a curvature
contribution). A numerical simulation, assuming the hedgehog form (\ref{hedgehog})
and finding the profile function $f(r)$ which minimizes $E$,
reveals that this is indeed what happens.
In particular, the energy $E/N$ of $N$-skyrmions in the range $3\leq N\leq20$
was computed for $\mu=1$, and the values are plotted in Figure 1. These values are
well-fitted by the curve $E/N=B+0.29/N^2$, where $B=5/3$ is the Bogomolny
bound.
\begin{figure}[htb]
\begin{center}
\includegraphics[scale=0.8]{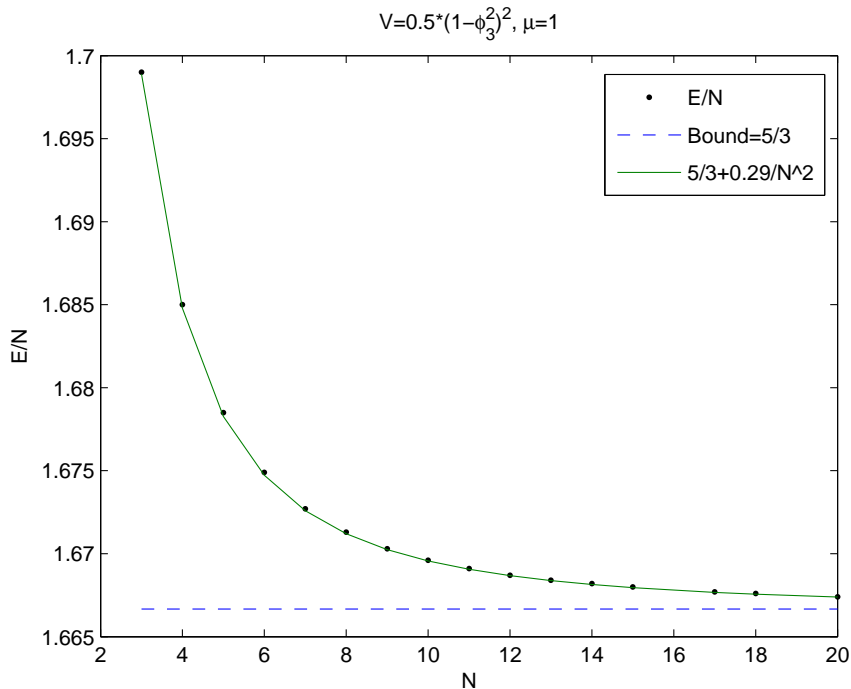}  
\caption{Energy of an $N$-Skyrmion in the
          system $V=\frac{1}{2}(1-\phi_3^2)^2$. \label{fig1}}
\end{center}
\end{figure}

Similar (although slightly less explicit) features occur for other `double-vacuum'
potentials, such as $V=\frac{1}{2}(1-\phi_3^2)$. In this case, one again
has a wall solution with a preferred length-per-charge $\beta$, which now
depends on the parameter $\mu$; below, we shall derive and plot the expression for 
$\beta=\beta(\mu)$. Isolated $N$-skyrmions in this system have previously
been studied \cite{weidig99}: they are rings with radii which grow
linearly with $N$, and $E/N$ is a decreasing function of $N$, similarly to
the previous example.

To calculate of $\beta(\mu)$, we assume homogeneity in the
$y$-direction, by taking the field to have the form
\begin{equation}
\label{NBS wall ansatz}
  \vec{\phi}=(\sin(f)\cos(\nu y),\,\sin(f)\sin(\nu y),\,\cos(f)).
\end{equation}
Here $\nu=2\pi/\beta$, and $f=f(x)$ is a function satisfying the boundary
conditions $f(x)\rightarrow 0$ as $x\rightarrow-\infty$ and
$f(x)\rightarrow\pi$ as $x\rightarrow\infty$. Note that this wall has $K=1$
(one unit of charge per period), and the energy per period is
\[
  E'=\frac{\beta}{4\pi}\int_{-\infty}^{\infty} \mathcal{E}\, dx,
\]
where
\[
   \mathcal{E}=\frac{1}{2}(f')^2+\frac{1}{2}(\nu^2+\mu)\sin^2 f
             +\frac{\mu\nu^2}{2}(f')^2\sin^2f.
\]
%
This functional $E'$ satisfies a Bogmolnyi-type inequality, namely
\begin{eqnarray}
\label{NBSwallEnergy}
  \frac{4\pi}{\beta}E' &=& \int_{-\infty}^{\infty} \frac{1}{2} \left(
  \sqrt{1+\mu\nu^2\sin^2f}f' - \sqrt{\mu+\nu^2}\sin f \right)^2\,dx\nonumber \\
        && + \sqrt{1+\mu\nu^2\sin^2f}\sqrt{\mu+\nu^2}\sin f f' \,dx\nonumber \\
  &\geq& \int_0^\pi \sqrt{1+\mu\nu^2\sin^2f}\sqrt{\mu+\nu^2}\sin f df,
\end{eqnarray}
with equality if and only if the Bogomolny equation
\begin{equation}
\label{NBSwallBog}
 f' = \frac{\sqrt{\mu+\nu^2}\sin f}{\sqrt{1+\mu\nu^2\sin^2f}},
\end{equation}
is satisfied. This equation can be integrated to give
\begin{eqnarray*}
\label{NBS wall profile function}
  x &=& x_0 + \frac{1}{\sqrt{\mu+\nu^2}} \left( \sqrt{\mu}\nu \mbox{arctan}\,
         \sqrt{\mu}\nu z + \frac{1}{2}\ln\frac{1+z}{1-z} \right), \\
  z &=& \frac{-\cos f}{\sqrt{1+\mu\nu^2\sin^2f}},
\end{eqnarray*}
and this implicitly determines the profile function $f(x)$. But without
knowing $f(x)$ explicitly, one can use (\ref{NBSwallEnergy}, \ref{NBSwallBog})
to obtain an expression for the minimal energy: it is
\begin{equation}
 E'=\frac{1}{2\nu}\sqrt{\mu+\nu^2} \left[ \left(\sqrt{\mu}\nu + 
    \frac{1}{\sqrt{\mu}\nu}\right) \mbox{arcsin}\,
     \sqrt{\frac{\mu\nu^2}{1+\mu\nu^2}} + 1 \right],
\end{equation}
and this is minimized with respect to variations in $\nu$ if
\begin{equation}
\label{NBSwallBeta}
  \left( \sqrt{\mu}\nu - \frac{1}{\sqrt{\mu}\nu} - 
    \frac{2\sqrt{\mu}}{\nu^3} \right) \mbox{arcsin}\,
      \sqrt{\frac{\mu\nu^2}{1+\mu\nu^2}} + 1 = 0.
\end{equation}
Solving (\ref{NBSwallBeta}) for $\nu$ then gives the preferred period
$\beta=2\pi/\nu$ as a function of $\mu$. This function $\beta(\mu)$ is plotted
in Figure 2, for the range $0\leq\mu\leq4$, from which one sees that the
dependence on $\mu$ is not very strong. Note that
$\beta/\pi\to2^{3/4}\approx1.68$ as $\mu\to\infty$.
%
\begin{figure}[htb]
\begin{center}
\includegraphics[scale=0.8]{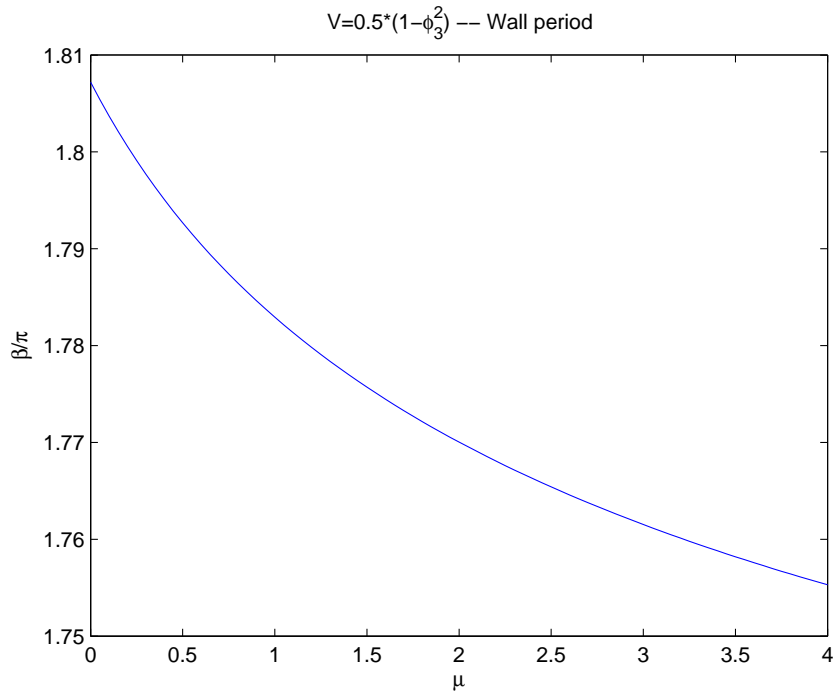}  
\caption{Preferred length-per-charge $\beta$ of a wall in the system
   $V=\frac{1}{2}(1-\phi_3^2)$.  \label{fig2}}
\end{center}
\end{figure}
%


\section{Skyrmion Chains}
\label{sk chains}

In this section, we consider fields which are anti-periodic
in $y$, and which satisfy the boundary condition
$\phi_3\to1$ as $x\to\pm\infty$. We call static solutions satisfying
these boundary conditions Skyrmion chains, and we shall investigate such
chains in the two systems $V=\frac{1}{2}(1-\phi_3^2)$ and $V=1-\phi_3$.

The energy $E'$ and magnetic charge $K$ per period are given by the same
expressions as before. It is not immediately obvious that $K$ has to be
an integer (since the field is anti-periodic rather than periodic),
but the following argument shows that it does (ie $K\in\mathbb{Z}$).
Suppose that $\vec\phi(x,y)$ satisfies the anti-periodicity condition
\begin{eqnarray*}
  (\phi_1(x,y+\beta), \phi_2(x,y+\beta), \phi_3(x,y+\beta))\\
    =(-\phi_1(x,y), -\phi_2(x,y), \phi_3(x,y)).
\end{eqnarray*}
%
We will construct a new field $\vec\psi(x,y)$ such that $K[\psi]=K[\phi]$, and
such that $\vec\psi$ is strictly periodic (with period $\beta$). But then as
described in the previous section, $K[\psi]$ is the degree of $\psi$, and
so is necessarily an integer; therefore $K[\phi]$ has to be an integer as well.
The new field is defined as $\psi(x,y)=R(y)\phi(x,y)$, where
\[
 R(y)=\left(\begin{array}{ccc}
  \cos(\pi y/\beta) &-\sin(\pi y/\beta) & 0 \\
  \sin(\pi y/\beta) & \cos(\pi y/\beta) &  0 \\
  0                 & 0                 & 1
  \end{array} \right).
\]
The charge density of $\psi$ is
\begin{eqnarray*}
 \mathcal{N}[R\phi] &=& \frac{1}{4\pi} (R\phi) \cdot (R\partial_1\phi)
                         \times (R\partial_2\phi +\partial_2R \phi) \\
                    &=& \mathcal{N}[\phi]
     + \frac{1}{4\pi} \phi \cdot \partial_1\phi\times [R^{-1}(\partial_2 R)\phi]\\
                    &=& \mathcal{N}[\phi]
     - \frac{1}{4\beta}\partial_1\phi_3.   
\end{eqnarray*}
And integrating this gives $K[R\phi]= K[\phi]$, as claimed, since
$\lim_{x\rightarrow\infty} \phi_3=\lim_{x\rightarrow-\infty} \phi_3$.

\subsection{Saturating the Bogomolny Bound}

The Bogomolny bound (\ref{bogomolny bound}) holds for skyrmion chains; as
was done previously for walls, one may write down simple configurations $W(z)$
which saturate the bound for potentials given by (\ref{sigma model potential}).
One possible choice, satisfying the chain boundary conditions, is
\begin{equation}
  W(z)= c\,\mbox{sech}(\pi z/\beta),
\end{equation}
where $c$ is a positive constant. The potential
defined by this field and (\ref{sigma model potential}) is
\begin{equation}
V = \frac{1}{2}\left(\frac{\pi}{\beta c}\right)^4(1-\phi_3)^2 
      \left[ \left( (c^2+1)\phi_3+(c^2-1) \right)^2 + 4 c^2\phi_2^2 \right].
\end{equation}
This potential $V$ has three vacua, at
\[
\vec{\phi} = (0,0,1),\,\left(  \frac{\pm2c}{1+c^2}, 0, \frac{1-c^2}{1+c^2} \right). \]
In the limit where $\beta\to\infty$, with $c\to0$ so that $\beta c$ remains
constant, we get the potential $V=(1-\phi_3)^4$. This was to be expected,
since it is exactly this potential which admits a `saturating' 1-skyrmion
solution on the plane \cite{piette&zakrzewski95}.

\subsection{Chains in the old baby Skyrme model}

In this subsection, we consider chains in the old baby Skyrme model, in other words
with potential $V=1-\phi_3$. We begin with an analytic
approximation describing a chain of well-separated skyrmions (this corresponds
to $\beta\gg1$), and then we describe some numerical results.

The $\beta\gg1$ approximation is an adaptation of the corresponding one for
widely-separated skyrmions in $\mathbb{R}^2$, used in \cite{piette&al95}.
It enables one to compute the
energy of a collection of skyrmions in terms of their separations and relative
phases. Let $W^{(1)}, W^{(2)}$ be two skyrmions (\ref{hedgehog}), with
phases $\chi^{(1)},\chi^{(2)}$, expressed in stereographic coordinates
(\ref{stereographic projection}). Suppose that these two skyrmions are
given a separation $D$, and consider the field $W=W^{(1)}+W^{(2)}$
representing their superposition.
In the limit where $D\rightarrow\infty$, the energy of $W$ is twice the energy
$E_1$ of a single skyrmion. One can calculate the leading contribution to the
difference
\[ I_{2} := E[W^{(1)}+W^{(2)}] - 2E_1, \]
in the limit where $D\rightarrow\infty$. The result \cite{piette&al95} is that
\begin{equation}
\label{Bessel1}
I_{2} \approx \frac{p^2\mu^2}{4\pi^2}
              \cos(\chi^{(2)}-\chi^{(1)})K_0(\sqrt{\mu}D)
\end{equation}
for large $D$, where $K_0$ is the modified Bessel function of order zero,
and where $p$ is a constant related to the decay of the profile
function $f(r)$ by 
\[ f(r) \approx \frac{p\mu}{2\pi} K_1(\sqrt{\mu}r). \]
This constant $p$ depends on $\mu$, and has to be determined numerically:
for example, when $\mu^2=0.1$, one finds $p=24.16$. This approximation is
called the dipole approximation, because the asymptotic formula for $I_2$
matches the
interaction energy of a pair of orthogonal scalar dipoles. Notice that the
sign of $I_2$ depends on the relative orientation $\chi^{(2)}-\chi^{(1)}$;
it follows that a pair of aligned skyrmions repel, while a pair of skyrmions
with relative phase $\pi$ attract.

This approximation can be adapted to the case of an infinite chain of skyrmions.
We shall do the computation both for in-phase skyrmions (so that the resulting
chain is periodic), and for anti-periodic chains where each pair of neighbouring
skyrmions has a relative phase of $\pi$. For each $j\in\mathbb{Z}$,
let $W^{(j)}$ be a skyrmion located at $(x,y)=(0,\beta j)$, with
phase $\chi=0$.  Then the fields
\[
  W^\pm := \sum_{j=-\infty}^\infty (\pm1)^j W^{(j)}
\]
are skyrmion chains, with relative phase $0$ or $\pi$ according to whether
one chooses the upper
or the lower sign. Let $I_\pm = E'[W^\pm]-E_1$ be the interaction energy per
period of the chain. One can show analytically that
\begin{equation}
\label{Bessel2}
  I_\pm \approx \sum_{j=1}^\infty (\pm1)^j \frac{p^2\mu^2}{4\pi^2}
      K_0(\sqrt{\mu} j\beta)
\end{equation}
to leading order, in the limit where $\beta\rightarrow\infty$. In fact, in
this limit, only the first term in the series (\ref{Bessel2}) is significant,
and so this approximation is really the same as (\ref{Bessel1}).
As before, $I_+(\beta)$ is a positive decreasing function, and
$I_-(\beta)$ is a negative increasing function (for large $\beta$).
So, a chain of well-separated skyrmions will extend if the skyrmions are
aligned, and contract if they are anti-aligned.

For the remainder of this subsection, we restrict to the anti-periodic case,
and in particular pose the question: is there a chain with a preferred
length-per-charge? In other words: which period $\beta$ (if any) will
minimize the energy $E'$? Anti-periodic chains in
this system were studied in \cite{gisiger96}
by using an ansatz for $\vec\phi$ which, in effect, imposed homogeneity
in the $y$-direction. Then there is a Bogomolny-type argument which yields
a formula for $E'$, and one can minimize $E'$ with respect to
$\beta$. The result is plotted in Figure 3 (dashed curve), for the
system with $\mu=1$.
%
\begin{figure}[htb]
\begin{center}
\includegraphics[scale=0.8]{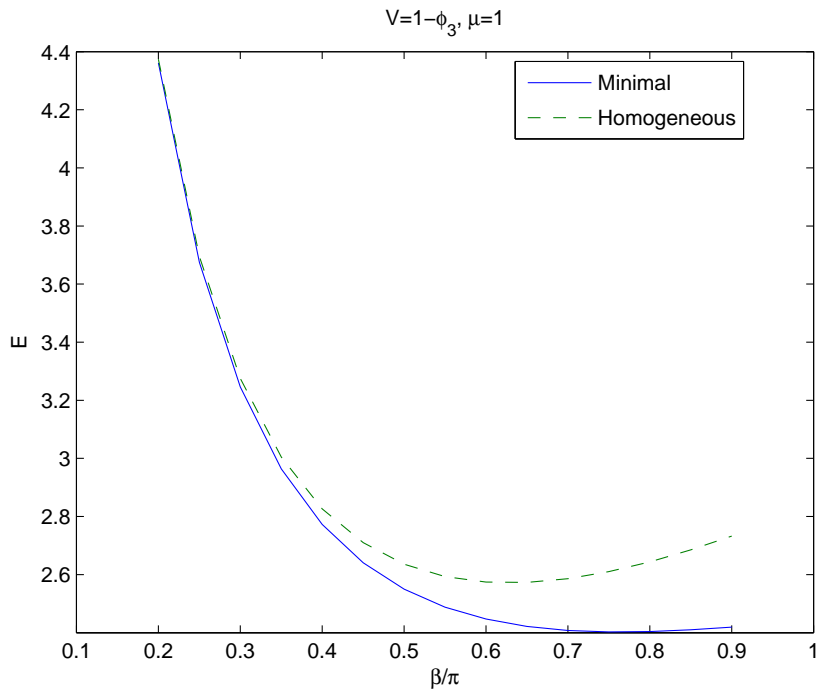}  
\caption{Energy-per-period of a chain in the system $V=1-\phi_3$, as
  a function of the period $\beta$. The solid curve gives gives the
  minimal energy; the dashed curve gives the energy of a chain which
  is $y$-homogeneous.  \label{fig3}}
\end{center}
\end{figure}
However, the actual solution, for each $\beta$, is not of
this homogeneous type, and the actual energy is lower than that of a homogeneous
field. This minimal energy has to  be determined numerically, and the result
of such a computation is also presented in Figure 3 (solid curve). 
The conclusion, therefore, is that a chain of
anti-aligned skyrmions in the old baby Skyrme system will contract until
its period $\beta$ reaches a value which minimizes $E'$; and for $\mu=1$
this preferred period is $\beta\approx0.76\pi$, as one sees from Figure 3.
Figure 4 depicts the chain at $\beta=0.76\pi$, with plots of the three
components $\phi_j$ as well as the energy density $\cal{E}$. 
%
\begin{figure}[htb]
\begin{center}
\includegraphics[scale=0.8]{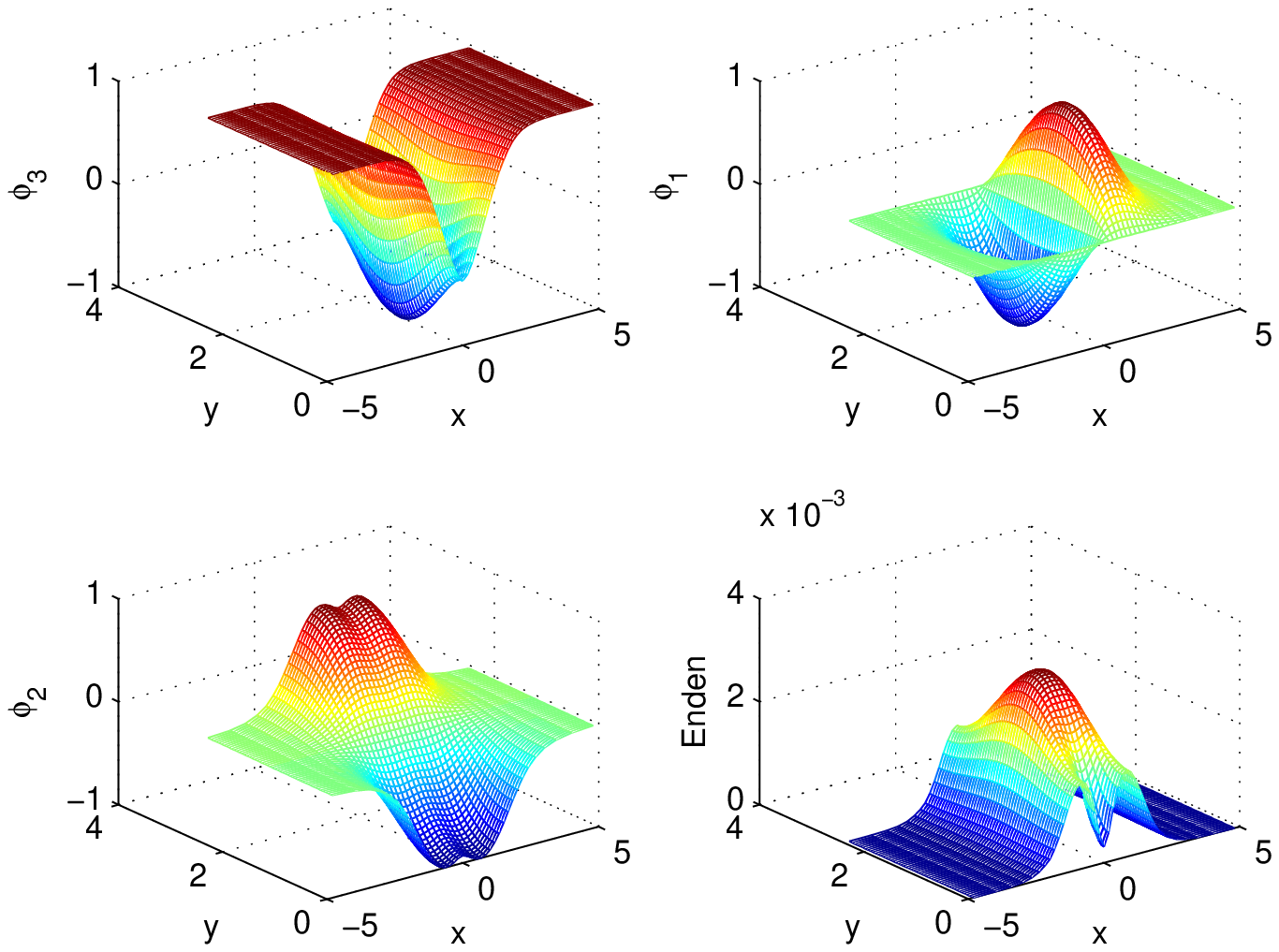}  
\caption{A chain in the system $V=1-\phi_3$, with $\mu=1$ and $\beta=0.76\pi$.
  The field components $\phi_j$, and the energy density $\cal{E}$, are plotted
  as functions of $x$ and $y$, over one period. \label{fig4}}
\end{center}
\end{figure}

The numerical procedure used in deriving Figures 3--5 is as follows.
The $xy$-space is replaced by a rectangular grid, with a transformation of
$x$ to ensure that $x\to\pm\infty$ is included. The expression for the energy
$E'$ is discretized with finite differences, and is then minimized using
a conjugate-gradient method. Richardson extrapolation is used to reduce the
error in $E'$ to significantly less than $1\%$.


\subsection{Chains in the new baby Skyrme model}

In this subsection, we consider chains in the new baby Skyrme model; that is,
with potential $V=\frac{1}{2}(1-\phi_3^2)$. The dipole approximation described
above works equally well in this case, and the same result is obtained:
namely that a chain of anti-aligned skyrmions will contract. As before, we
pose the question of whether there is a preferred period $\beta$ corresponding
to a minimal-energy chain.

A new feature in this case is that the system also admits wall solutions.
One may construct a field with chain boundary conditions by taking it to
consist of a parallel wall-antiwall pair. From this it is clear that there
is now a possible decay mode for chains: namely that a chain could
split into a wall and an antiwall which separate, and move in the negative
$x$-direction and the positive $x$-direction respectively. Whether or not
the wall and antiwall actually do move apart depends on the force between
them, and this force can be studied analytically, as follows.

Let $\phi^{(1)}$ be a domain wall (\ref{NBS wall ansatz}) with
$\nu^{(1)}=\pi/\beta$, with profile function $f^{(1)}$ given by
(\ref{NBS wall profile function}), and located at $x=-D/2$ for some
real number $D>0$.  Let $\phi^{(2)}$ be a second domain wall given by
(\ref{NBS wall ansatz}) with $\nu^{(2)}=-\pi/\beta$, with profile function
$f^{(2)}(x) = f^{(1)}(-x)$. So this second wall is an antiwall, and it is
located at $x=D/2$. Let $W^{(1)}$ and $W^{(2)}$ be the stereographic
projections of these, as in (\ref{stereographic projection}). Finally,
define a superposition $W$ by
\[
  \frac{1}{W} = \frac{1}{W^{(1)}} + \frac{1}{W^{(2)}}.
  \]
The field $W$ satisfies the boundary conditions for anti-periodic chain with
period $\beta$, and the charge per unit period is 1. This field resembles
a parallel wall-antiwall pair, separated by a distance $D$, with $\phi^3=-1$
between the walls and $\phi_3=1$ on either side.
We want the interaction energy
\[
   I_w := E'[W] - E'[W^{(1)}] - E'[W^{(2)}]
\]
of the superposition, which is a function of $\beta$ and $\mu$.

Now $I_w\rightarrow0$ as $\beta\rightarrow\infty$, and we have calculated
the leading contribution to $I_w$ in this limit.  The calculation is similar
to that for two isolated skyrmions in \cite{piette&al95}, but is more
complicated in practice, for the following reason. In the case of two
isolated skyrmions, the leading contribution is found using an elegant
argument based
on the Euler-Lagrange equations; in the case of two walls, however, the
analogous term vanishes, and so we must work to higher order. We find that
\begin{eqnarray*}
I_w &\approx& \frac{4\mu\exp(4\sqrt{\mu}\nu
                \mbox{arctan}\,\sqrt{\mu}\nu)}{\nu(1+\mu\nu^2)^2}\\
           & &\times (2\nu^4+2\mu\nu^2-1 )D\exp(-2D\sqrt{\mu+\nu^2})
\end{eqnarray*}
provided $D\gg1$, where we have written $\nu=\pi/\beta$ .
So a pair of well-separated domain walls will attract when
$\beta^4 - 2\mu\pi^2\beta^2 - 2\pi^4  > 0$ and repel otherwise.
Therefore the approximate prediction is that a chain is stable with respect
to separation into two walls provided that its period $\beta$ is greater
than the critical value
\begin{equation}
\label{criticalperiod}
   \beta_c = \pi \sqrt{\mu+\sqrt{\mu^2+2}}.
\end{equation}
But for $\beta < \beta_c$ chains will decay into separating wall-antiwall pairs.

Recall from section 3 that each wall in a well-separated wall-antiwall pair
has a preferred
period $\beta$: it is obtained by solving (\ref{NBSwallBeta}) for $\nu$, and
then setting $\beta=\pi/\nu$. (We take $\beta$ to have half the value it had in
section 3, so that the charge-per-period of the wall-antiwall system is $K=1$.
As a consequence, the chain is anti-periodic rather than periodic.)
From Figure 2 we see that for $\mu=1$, the preferred period is $\beta=0.89\pi$.
By contrast, (\ref{criticalperiod}) gives $\beta_c=1.65\pi$. So if we begin with
a wall-antiwall pair with large $\beta$, then each wall in the pair will continue
to contract until the period reaches $\beta_c$, whereupon they will begin to repel
each other.

A numerical simulation for the case $\mu=1$ reveals that this is indeed what
happens, with the critical value of the period being only $10\%$ below that
predicted by (\ref{criticalperiod}). This result is illustrated in Figure 5,
which plots the energy $E'$ of a chain, numerically-minimized for a range of
(fixed) values of $\beta$.
%
\begin{figure}[htb]
\begin{center}
\includegraphics[scale=0.8]{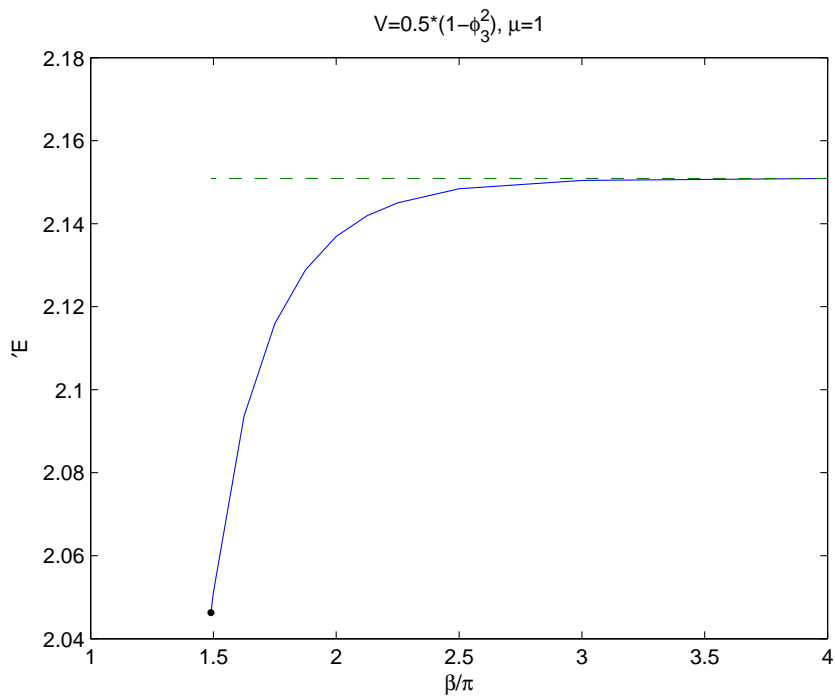}  
\caption{Energy versus $\beta/\pi$ of a chain in the system
   $V=\frac{1}{2}(1-\phi_3^2)$. For $\beta<1.49\pi$, the chain is
   unstable. \label{fig5}}
\end{center}
\end{figure}
The dashed line is the energy $E=2.151$ of an
isolated skyrmion in this system, and so one sees that for $\beta\geq4\pi$
each skyrmion in the chain hardly notices its neighbours. Energy can be
lowered by reducing the period (the chain tends to contract); but once the
period reaches $\beta\approx1.49\pi$, the separating-wall instability sets in.
So for chains in this system there is no preferred length-per-charge.

\section{Conclusions}

In single-vacuum systems such as $V=1-\phi_3$, one has chains with a
preferred length-per-charge. They are stable against periodic perturbations;
their stability against more general perturbations has not been investigated
here.
In double-vacuum systems such as $V=\frac{1}{2}(1-\phi_3^2)$, one has
stable wall solutions, but there are no stable chains. The walls are,
in some sense, the primary objects in such systems --- skyrmions are
made out of them (although one can, in certain cases, also have crystalline
chunks \cite{ward04}). The same is true one dimension higher up, namely
for the Skyrme model. Skyrme walls were discussed in \cite{battye&sutcliffe98},
and skyrmions are hollow shells constructed out of this wall material \cite{BS02}.
Introducing a potential can change things: for example, potentials of
$1-\phi_3$ type \cite{BS06}, or of more general type \cite{KPZ06}.

It would be interesting to extend the investigations described here to other,
related, systems. For example, in 2+1 dimensions, the baby Skyrme model is
related to abelian Higgs models which admit local and semi-local vortices
\cite{W02}; what are the properties of chains and walls in this general family
of systems, and their $\mathbb{CP}^n$ generalizations?
Going to 3+1 dimensions, do stable chains exist in the Skyrme model, and how does
the choice of potential affect such structures?

\begin{acknowledgments}
This work was supported by a studentship, and research grants
``Strongly Coupled Phenomena'' and ``Classical Lattice Field Theory'',
from the UK Science and Technology Facilities Council.
\end{acknowledgments}

\end{document}